\documentclass{elsart}
\usepackage{natbib}
\usepackage{graphicx}
\usepackage{bbm}
\usepackage{psfrag}
\usepackage{listings} 
\usepackage{url}
\bibpunct{[}{]}{,}{a}{}{;}
\begin{document}
\lstset{language=C}

\begin{frontmatter}
\title{A program generating homogeneous random graphs with given weights}
\author[Bogacz]{L. Bogacz},
\author[Burda]{Z. Burda},
\author[Bogacz]{W. Janke},
\author[Burda]{B. Waclaw}
\address[Bogacz]{Institut f\"ur Theoretische Physik, Universit\"at Leipzig, Augustusplatz 10/11, \\ 04109 Leipzig, Germany}
\address[Burda]{Institute of Physics, Jagellonian University, Reymonta 4, 30-059 Krakow, Poland}

\begin{abstract}
We present a program package which generates homogeneous random graphs with probabilities prescribed by the user. The statistical weight of a labeled graph $\alpha$ is given in the form $W(\alpha)=\prod_{i=1}^N p(q_i)$, where $p(q)$ is an arbitrary user function and $q_i$ are the degrees of the graph nodes. The program can be used to generate two types of graphs (simple graphs and pseudo-graphs) from three types of ensembles (micro-canonical, canonical and grand-canonical).\\
PACS: 89.65.-s, 89.75.Fb, 89.75.Hc \\
Keywords: random graphs, complex networks, Markov process, Monte Carlo method\\
\end{abstract}
\end{frontmatter}

\section{Program summary}
\emph{Title of the program:} GraphGen. \\
\emph{Catalogue identifier:} \\
\emph{Program obtainable from:} \\
\url{http://www.physik.uni-leipzig.de/~bogacz/graphgen/} \\
\emph{Computer for which the program is designed and others on which it has been tested:} 
PC, Alpha workstation. \\
\emph{Operating systems or monitors under which the program has been tested:} Linux, Unix, MS Windows XP.\\
\emph{Programing language used:} C. \\
\emph{Memory required to execute with typical data:} 300k words for a graph with 1000 nodes and up to 50000 links. \\
\emph{No. of bits in a word:} 32. \\
\emph{No. of processor used:} 1. \\
\emph{Has the code been vectorized or parallelized:} No. \\
\emph{No. of bytes in distributed program, including test data etc.:} 15k.\\
\emph{Distribution format:} Compressed tar file. \\
\emph{Keywords:} random graphs, complex networks, Markov process, Monte Carlo method.\\
\emph{Nature of the problem:} The program generates random graphs. The probabilities of graph occurrence are proportional to their statistical weight, dependent on node degrees defined by arbitrary distributions.\\
\emph{Method of solution:} The starting graph is taken arbitrary and then a sequence of graphs is generated. Each graph is obtained from the previous one by means of a simple modification. The probability of accepting or rejecting the new graph results from a detailed balance condition realized as Metropolis algorithm. When the length of the generated Markov chain increases, the probabilities of graph occurrence approach the stationary distribution given by the user-defined weights ascribed to the graphs.\\
\emph{Restrictions on the complexity of the problem:} None.\\
\emph{Typical running time:} Less than two minutes to generate $10^5$ graphs of size $10000$ nodes and $30000$ links on a typical PC. \\
\emph{Unusual features of the program:} None. \\

\section{Introduction}
Complex networks can be easily found in the real world. 
If the world objects are represented by nodes, and the interactions between them by edges then phone calls, computer connections, disease spread diagrams and human contacts are only a few examples of such networks. The recent improvements of computer technology has made the data acquisition easier and in consequence has led
to a development of large databases of topology of observed networks. It turned out that completely independent networks often share common features, such as small world effect, fat tail in node degree distribution or large clustering. These effects caused that random graph theory, being mainly studied
by pure mathematics so far, has attracted the attention of physicists and other natural sciences (for reviews see \cite{ref:ab,ref:dm,ref:n1}).

There are two natural approaches to simulate networks as random graphs: diachronic \cite{ref:ab,ref:dm,ref:n1} and synchronic \cite{ref:bck,ref:bl1,ref:dms,ref:bjk1,ref:fdpv,ref:pn1,ref:kahng,ref:bbw}. In the first the network evolution in time is being investigated. One simulates the process of growth and checks how different mechanisms influence the emerging final graphs. In the latter approach a statistical ensemble of graphs is constructed and methods of statistical mechanics are applied. Each graph has a weight determining the probability of its occurrence during random sampling. 
The emergence of real networks usually is a complex process and computer simulations require the application of both approaches together. For example, the Internet is still a growing network, but its older parts also evolve.

The program package we describe in this paper uses the synchronic approach, which is a natural extension of Erd\"os and R\'enyi ideas \cite{ref:er,ref:bb}. A statistical ensemble of graphs is built by assigning a weight to each labeled graph in the given set of graphs. The weights can be chosen arbitrary. 
As an illustration, in the program we chose them to depend on degrees of individual nodes.
If more complicated weights are needed, slight modifications of the program source code are required.

The program package can be used to generate graphs which mimic scale-free networks, i.e., networks with power-law degree distribution \cite{ref:bamodel} or, in general, any desired degree distribution. One crucial point must be explained here. For finite graph size all the methods devoted to generating fat-tailed degree distributions introduce a cut-off effect, since it is a property of networks itself \cite{ref:bk,ref:cutoff,ref:cutoff2}. The same is true for the program presented in this paper and there is no way to omit this effect without changing other properties such as lack of self- and multiple-connections.

The rest of this paper is organized as follows. In section \ref{definitions} we present definitions of graphs, statistical ensembles of graphs and partition functions of the models presented in the paper. Section \ref{methods} contains the description of the method used to generate graphs. In the final section, we outline the program compilation and usage.

\section{\label{definitions}Definitions}
\begin{figure}[h]
\begin{center}
\includegraphics[height=4cm]{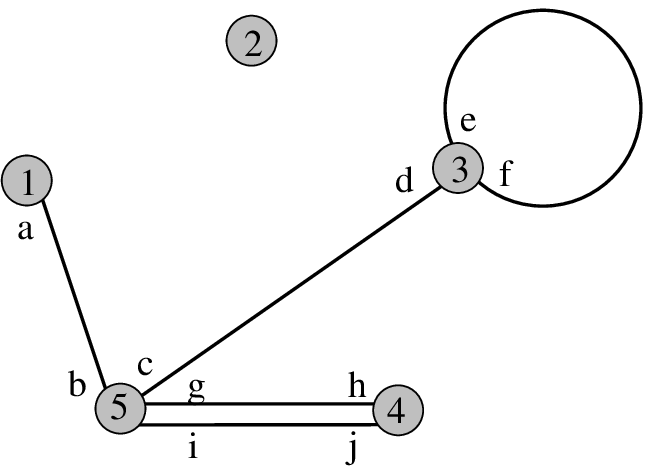} 
\caption{An example of a graph with $N=5$ nodes and $L=5$ links. Positions of vertices in the picture are meaningless. The only information which matters is connectivity.\label{example_graph}}
\end{center}
\end{figure}

Let us start with basic definitions. A graph is a set of $N$ nodes (vertices) connected by $L$ edges (links) (see the example in fig. \ref{example_graph}). The edges can be directed or undirected, but in this paper we constrain ourselves only to graphs with undirected edges. The graphs without multiple connected or self-connected nodes will be called simple graphs (or graphs). Graphs containing multiple-connected or self-connected nodes will be denoted pseudographs or degenerated graphs. A graph does not need to be connected.

Both simple graphs and pseudographs can be represented by an adjacency matrix. For graphs with $N$ nodes this is a $N\times N$ matrix with elements $A_{ij}$ equal to the number of edges connecting node $i$ and node $j$ (for $i \neq j$). Diagonal elements $A_{ii}$ count twice the number of self-connecting edges attached to node $i$,
because we count each endpoint of link once.
For example, the adjacency matrix of the graph shown in fig. \ref{example_graph} has the following form:
\begin{equation}
\renewcommand*{\arraystretch}{0.8}
\mathbf{A}=\left(\begin{array}{ccccc}
 0 & 0 & 0 & 0 & 1\\
 0 & 0 & 0 & 0 & 0\\
 0 & 0 & 2 & 0 & 1\\
 0 & 0 & 0 & 0 & 2\\
 1 & 0 & 1 & 2 & 0\\
\end{array}\right) \, .
\end{equation}
The adjacency matrix is symmetric for any type of graph with undirected edges. Additionally, for simple graphs all diagonal elements are zero, and all other elements are zero or one. The sum of elements in the $i$-th row (or in the $i$-th column) gives the degree (the order) of the vertex $i$, i.e., the number of edges connected to that vertex.

\begin{figure}[h]
\begin{center}
\includegraphics[height=5.0cm]{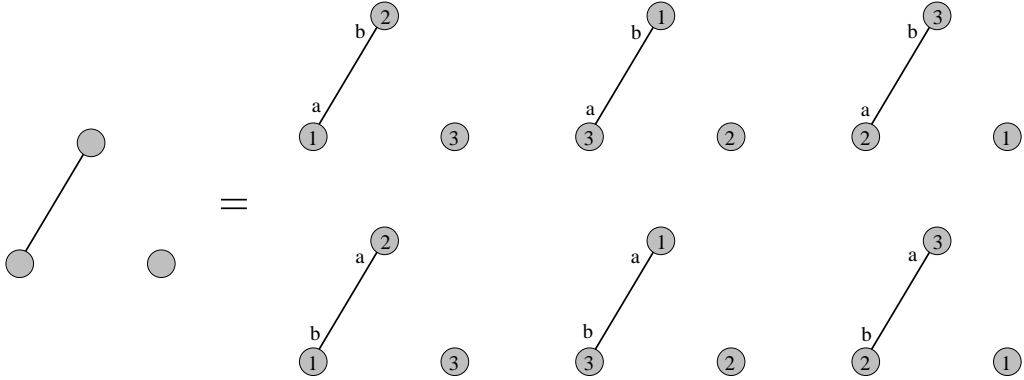} 
\caption{The unlabeled graph on the left corresponds to the six possible labeled graphs on the right.\label{labels}}
\end{center}
\end{figure}

A statistical ensemble of graphs is defined by ascribing a statistical weight to every graph in a given set. Among many possible choices we have defined and implemented in the program three sets of graphs \cite{ref:bbw}:
\begin{enumerate}
\item The {\em canonical} ensemble consists of all labeled graphs with fixed number of nodes $N$ and edges $L$. It is a generalization of well known Erd\"os and R\'enyi graphs, where one connects $N$ nodes by $L$ edges chosen at random from all possibilities.
\item The {\em grand-canonical} ensemble is the ensemble of all labeled graphs with fixed number of nodes $N$. The number of edges $L$ is varying. This is a generalization of the so-called binomial model, also introduced by Erd\"os and R\'enyi \cite{ref:er}.
\item The {\em micro-canonical} ensemble consists of all labeled graphs with fixed degree of nodes given by a set of numbers ${q_1,\dots, q_N}$.
\end{enumerate}
Each of those ensembles can exist in two versions, consisting of only simple graphs or also pseudographs. By fully labeled (called for simplicity labeled) graph we mean a graph with labeled nodes and edges. Each edge has two labels, attached to two endpoints (see fig. \ref{example_graph}).
Graphs which are identical in the sense of shape are not necessarily identically labeled graphs. 
Consider for example the graph shown in fig. \ref{labels}. This unlabeled graph has 6 different realizations as labeled graph, shown on the right hand side of fig. \ref{labels}. 

The weight of each graph in the ensemble is defined in two steps. 
First we introduce a uniform configurational weight $1/(N!(2L)!)$ for each labeled graph. 
This weight is to compensate the number of permutations of indices.
However, we will see below that for graphs possessing special symmetries the number of distinct labeled
graphs is smaller than $N!(2L)!$ and therefore some additional factor remains.
The partition function for the canonical ensemble of graphs with uniform measure is defined as:
\begin{equation}
Z_u(N,L) = \sum_{\alpha'\in flg(N,L)} \frac{1}{N!(2L)!} = \sum_{\alpha\in g(N,L)} w(\alpha), 
\end{equation}
where $flg(N,L)$ is the set of all fully labeled graphs with given number of nodes $N$ and edges $L$
and $g(N,L)$ denotes the set of all unlabeled graphs of size $N$ and $L$. The weight function $w(\alpha)$ is defined in such a way that $w(\alpha)N!(2L)!$ is the total number of fully labeled graphs corresponding to the unlabeled graph $\alpha$.
If one considers only simple graphs the edge labeling can be abandoned. In this case the edge position is uniquely determined by two nodes at its endpoints. 
The $1/(2L)!$ factor cancels all possible edge relabelings, so exactly the same model can be defined when one replaces the uniform measure $1/(N!(2L)!)$ by $1/N!$ and does not introduce edge labels. 
The $Z_u(N,L)$ function defined above is just the partition function of the Erd\"os-R\'enyi model \cite{ref:er}.

On the basis of the canonical partition function $Z_u(N,L)$, we define the partition function for the grand-canonical ensemble \cite{ref:bbw}:
\begin{equation}
Z_u(N,\mu) = \sum_L \exp (-\mu L) Z_u(N,L) = \sum_L \exp (-\mu L) \sum_{\alpha\in g(N,L)} w(\alpha),
\end{equation}
where $\mu$ can be interpreted as chemical potential for edges. Defining $p$ as $\frac{p}{1-p}\equiv \exp(-\mu)$ one
realizes that $Z_u(N,\mu)$ is the partition function for the binomial model.
The partition function for the micro-canonical ensemble with given node degree sequence ${q_1,q_2,...,q_N}$ can be defined as:
\begin{equation}
Z_u(N,\left\{ q_i \right\}) = \sum_{\alpha\in g(N,L)} \left( \prod_{i=1}^N \delta \left( q_i(\alpha) - q_i\right) \right) \, w(\alpha) \, ,
\end{equation}
where $\delta(x)=1$ if $x=0$ and zero otherwise. The $q_i(\alpha)$ gives the degree of the $i$-th vertex of graph $\alpha$. Consider as an example the canonical ensemble of graphs with $N=3$ nodes and $L=2$ edges. There are 6 possible unlabeled graphs, shown in table \ref{tab1}. For each graph the number of corresponding labeled graphs, the uniform weight and the normalized occurrence probabilities $p(\alpha)=w(\alpha)/\sum_\beta w(\beta)$ are also shown.

The uniform weight ($w(\alpha)=1, \forall\alpha$) leads to networks with Poissonian degree distribution.
In the real world one rather observes networks with fat tails. 
Therefore we introduce an additional functional weight $W(\alpha)$, which is defined as
\begin{equation}
W(\alpha)=\prod_{i=1}^N p(q_i),
\label{graph_weight}
\end{equation}
where the $p(q_i)$ function depends on the degree $q_i$ of $i$-th graph's vertex.
The $p(q)$ can be chosen to obtain desired properties of the statistical ensemble.
For example, one can show (see, e.g., \cite{ref:dms}) that for the canonical ensemble of graphs the choice $p(q)=q! \pi(q)$ leads to the average degree distribution $\pi(q)$ in the limit $N \rightarrow \infty$. 
Therefore, taking $\pi(q)\propto q^{-\gamma}$ we obtain scale-free networks.

The partition functions for canonical, grand-canonical and micro-canonical ensembles with additional weight $W(\alpha)$ are:
\begin{equation}
Z(N,L) = \sum_{\alpha'\in flg(N,L)} \frac{W(\alpha')}{N!(2L)!} = \sum_{\alpha\in g(N,L)} w(\alpha)W(\alpha),
\end{equation}
\begin{equation}
Z(N,\mu) = \sum_L \exp (-\mu L) \sum_{\alpha\in g(N,L)} w(\alpha) W(\alpha),
\end{equation}
\begin{equation}
Z(N,\left\{ q_i \right\}) = \sum_{\alpha\in g(N,L)} \left( \prod_{i=1}^N \delta \left( q_i(\alpha) - q_i\right) \right) \, w(\alpha) W(\alpha) \, .
\end{equation}
Because of the chosen form, the functional weight $W(\alpha)$ has the same value for each graph taken from the micro-canonical ensemble. Thus it factorizes and has no influence on properties of the micro-canonical ensemble. However, in the general case when one defines a more complicated function $W(\alpha)$, for example dependent on the number of certain motives present in the graph, it will modify the relative weights of graphs also in the micro-canonical ensemble. To introduce such a function, modifications of the program code are required.

\section{\label{methods}Methods}
The purpose of the presented program is to generate graphs with probabilities proportional to their statistical weights. Unfortunately there is no efficient algorithm which would be able to pick up an element from a large set with given probability. The naive algorithm which would pick up a random element and then accept or reject it with probability proportional to its weight would be very inefficient because of low acceptance rates. Therefore we use instead a Markov chain Monte Carlo technique, known from simulations of physical systems \cite{ref:mc1,ref:mc2}. We construct a guided random walk in the configuration space of graphs.
In each step, the program recursively generates a new graph $\alpha_{t+1}$ by modification of the current one $\alpha_t$. 
In this way we obtain a Markov chain of configurations $\alpha_0 \rightarrow \alpha_1 \rightarrow \alpha_2 \rightarrow \dots\;$. The chain is determined by the transition probabilities matrix $P(\alpha \rightarrow \beta)$ encoding how the modification of the graph $\alpha$ will lead to graph $\beta$, and the initial configuration.
If the process is ergodic (which roughly means that all configurations are accessible) 
and if the probabilities fulfill the detailed balance condition:
\begin{equation}
W(\alpha) P(\alpha \rightarrow \beta) = W(\beta)P(\beta \rightarrow \alpha),
\label{balance}
\end{equation}
where $W({\alpha})$ is the weight of graph $\alpha$, then the frequencies of graph occurrence approach the distribution $W(\alpha)/Z$ as the number of steps goes to infinity. 
In the program presented here, $P(\alpha \rightarrow \beta)$ is chosen as:
\begin{equation}
P(\alpha \rightarrow \beta)=\min\left\{1,\frac{W(\beta)}{W(\alpha)}\right\},
\label{metropolis}
\end{equation}
which is known as Metropolis algorithm \cite{ref:metrop}.
Depending on the considered graph ensemble we propose as elementary move one of the three transformations described below.

\begin{figure}[h]
\begin{center}
\psfrag{a}{\footnotesize T}
\psfrag{b}{\footnotesize add/remove}
\psfrag{c}{\footnotesize X}
\includegraphics[height=6.5cm]{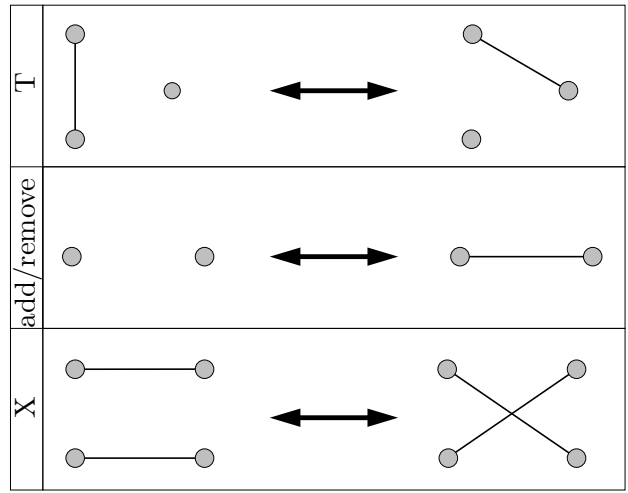} 
\caption{Three types of graph modification used for generating graphs from canonical (T-move), grand-canonical (add/remove) and micro-canonical (X-move) ensembles.\label{moves}}
\end{center}
\end{figure}

The first graph transformation called ``T-move'' is used to modify graphs belonging to the canonical ensemble. First, one node $j$ and one edge $i\rightarrow k$ are chosen at random. 
Then we rewire the edge to $i\rightarrow j$ which means that the edge is detached from its endpoint $k$ and attached to $j$ (see fig. \ref{moves}). The total number of edges $L$ is thus conserved but
the degrees of the vertices $k$ and $j$ are changed: $q_k\rightarrow q_k-1, q_j\rightarrow q_j+1$.
The probability for accepting the transformation is given by formula (\ref{metropolis}) as
\begin{equation}
P_a(\alpha\rightarrow\beta) = \min\left\{1,\frac{p(q_k-1)p(q_j+1)}{p(q_k)p(q_j)}\right\},
\label{prob_can}
\end{equation}
where we explicitly used the form of the functional weight given by (\ref{graph_weight}).

The second graph transformation which we consider is used to modify graphs belonging to the grand-canonical ensemble. For this ensemble we introduce two reciprocal transformations -- addition and deletion of a link. Both of them preserve the number of nodes in the graph but change the number of edges (see fig. \ref{moves}). 
The decision which of those two is used in each elementary step is taken at random with probability $1/2$. 

As it was shown in \cite{ref:bbw}, the probabilities of accepting addition and removal of a link
are respectively:
\begin{equation}
P_a(\alpha\rightarrow\beta) =
\min\left\{1, \exp(-\mu) \,
\frac{N^2}{2(L(\alpha)+1)} \, \frac{W(\beta)}{W(\alpha)} \right\} ,
\label{prob_grand1}
\end{equation}
and
\begin{equation}
P_a(\beta\rightarrow\alpha) =
\min\left\{1, \exp(+\mu)  \,
\frac{2L(\beta)}{N^2} \, \frac{W(\alpha)}{W(\beta)}\right\}.
\label{prob_grand2}
\end{equation}

The last transformation called ``X-move'' is used to modify graphs from the micro-canonical ensemble \cite{ref:maslov}. 
First, two links $l_1$, $l_2$ are chosen randomly from all existing edges. 
Assume that $l_1$ connects vertices $i_{a}$, $i_{b}$ and $l_2$ connects $j_{a}$, $j_{b}$. 
Next we exchange their endpoints so that $l_1,l_2$ point onto $j_{b},i_{b}$, respectively. 
The degrees of all four nodes remain unchanged (see fig. \ref{moves}). 
The probability of  accepting the move is equal to one, because the weights of all labeled graphs in the micro-canonical ensemble are identical.

If we want to generate only simple graphs, additional constraints must be introduced:
we reject all moves leading to self- or multiple-connections.
This does not change the probabilities of graph occurrences but only restricts the
configuration space to what we need.

Because of the chosen graph generation method, each simulation should start from a ``thermalization'' sequence. Graphs generated during this sequence are not saved and no measurements are made. This is necessary for the graph occurrence probabilities to approach the proper distribution resulting from the weight function since we usually start
from a graph which does not need to be ``typical'' in the given ensemble.
The length of the ``thermalization'' sequence depends on the chosen ensemble, graph size and weight function. 
To estimate this length one may look at one particular property of a graph like degree distribution
and check how many steps are needed to obtain the expected shape,
using $\chi^2$ function calculated for theoretical and measured degree distribution.
Starting from one particular configuration, e.g., a Poissonian random graph, one has to wait until $\chi^2\approx 1$.
One can use the {\em degdist} program, included in the package, to generate node degree distributions for different lengths of thermalization sequence. 
Comparing those with theoretical distributions and calculating $\chi^2$ one may 
find an appropriate length of ``thermalization'' sequence.

The graphs generated by the program are correlated. The autocorrelation time depends on program parameters but also on the measured observable. As an example we report the autocorrelation time for the average clustering coefficient and for the total number of triangles in the graphs generated from the canonical ensemble. The autocorrelation time for graphs with unit weight, with $N=100$ nodes and $L=1000$ links, when a sweep contains $100$ graph modification trials (see the SWEEP definition in the next section) is $t_{ac}\approx 3.9$ for the clustering coefficient and $t_{ac} \approx 4.9$ for the number of triangles. The correlation length grows approximately linearly with the number of graph links. To reduce this autocorelation time simply increase the SWEEP parameter value.

\section{\label{program_description}Program description}
\subsection{Source Code}
We provide two programs for the generation of the described graph ensembles. Both of them are written in the ``C'' language. The first, {\em graphgen} is designed for generating graphs and saving them to a file. The user can make desired operations on the generated and saved sample.
The second program called {\em degdist} demonstrates how to write a simple program calculating
some quantities like the average degree distribution without saving the intermediate results to a file. Both programs use the same procedures, collected in a few separate files.
The complete set of source files is presented below:

\begin{enumerate}
\item {\tt init.c} -- set of functions used to build (initialize) a new graph. The initial graph is constructed by adding some links between randomly chosen nodes.
\item {\tt links.c} -- functions used to perform operations on graphs. These are for example inserting or removing a link from a graph, choosing links or edges at random etc.
\item {\tt sweep.c} -- functions performing three types of graph modification (T-move, addition/removal of links, X-move) used to modify the graphs from all ensembles.
\item {\tt save\_load.c} -- functions used for loading the initial graph from a file and saving generated graphs.
\item {\tt graphgen.c} -- main function of program {\em graphgen}, responsible for reading parameters from the command line and management of the graph generation process.
\item {\tt degdist.c} -- the program {\em degdist} that generates the histogram of degree distribution for a given ensemble of graphs.
\end{enumerate}

First we describe the program {\em graphgen}. 
The source code has been divided into eight files: three header files ({\tt def.h, functions.h, variables.h}) and five source code files (the above 1-5). 
The {\tt def.h} file should be edited before compilation. Constants defined therein determine the ensemble type used for the simulation, the weight function, the save and load file format and limits for the maximal number of nodes and edges. The complete list of options will be described in detail in subsection \ref{compilation}.
The execution and description of output data file is given in subsections \ref{exec} and \ref{save_load_format}.

The program {\em degdist} is described in subsections \ref{deg_comp} and \ref{deg_exec}.
The first one is devoted to compilation while the last one gives some informations about execution and output format.

\subsection{\label{compilation}Compilation of graphgen}

To increase program efficiency, the decision which type of ensemble is going to be simulated is made before program compilation. Therefore before program compilation one should check and modify the definitions in the {\tt def.h} file if necessary. The structure of the file corresponds to the definitions of macro constants in the ``C'' language. Each line has the following form:
\begin{lstlisting}
#define NAME value
\end{lstlisting}
where NAME and value can be any pair from the list:
\begin{itemize}
\item  ENSEMBLE \ \ [1, 2, or 3]: This value determines what type of ensemble the program uses
to generate graphs. Use $1$ for micro-canonical, $2$ for canonical, and $3$ for grand-canonical  ensemble.
\item  GRAPH\_TYPE \ \ [1, 2, or 3]: This determines if self- and multiple-connections are allowed. Use $1$ to generate simple graphs only, $2$ to generate multi-graphs with multiple-connections but without self-connections, and $3$ to generate pseudographs with self- and multiple-connections.
\item  SAVE\_FORMAT \ \ [1, 2, or 3]: This constant sets the default format for saving and loading a graph. Use $1$ for full adjacency matrix format, $2$ for short adjacency matrix format, and $3$ for node order format (for a detailed description, see subsection \ref{save_load_format}).
\item  WEIGHT\_FUNCTION \ \ $p(q)$: The function $p(q)$ determines the contribution from one of the nodes to the total graph weight (\ref{graph_weight}). Here $q$ is an integer number equal to the node degree. The function $p(q)$ can be defined in any format consistent with the ``C'' language (for example $1.0/q$).
It is used only if canonical or grand-canonical ensembles are chosen and the parameter RATIO\_WEIGHT\_FUNCTION is not defined.
\item  RATIO\_WEIGHT\_FUNCTION $p(q+1)/p(q)$: In the calculation of transition probabilities (\ref{prob_can}), (\ref{prob_grand1}), (\ref{prob_grand2}) only the ratio $p(q+1)/p(q)$ is used. Therefore it is better to define this ratio instead of the function $p(q)$. This reduces round-off errors and increases efficiency of the program (for example use $q+1$, when $p(q)=q!$, which avoids calculating the factorial). If the RATIO\_WEIGHT\_FUNCTION is defined then the WEIGHT\_FUNCTION is ignored. The ratio can be defined in any format consistent with the ``C'' language.
\item  NV [integer number]: This sets the upper limit for the number of graph vertices and restricts the size of the graph to be generated or loaded. The larger the limit is, the more memory is required to run the program.
\item  NL [integer number]: As NV but for graph edges.
\item  SWEEP [integer number]: To obtain a new graph from the previous one, the program modifies the graph by a sequence of elementary transformations described in section \ref{methods}. The parameter SWEEP denotes the number of attempts of such elementary transformations.
\item  THERM [integer number]: This value determines the number of sweeps to be made at the beginning of a simulation without saving the generated graphs. Such starting sequence is necessary to ``thermalize'' the system.
\item  GRAPHS [integer number]: Determines how many graphs should be generated (saved or printed). After the starting sequence, the generated graphs are saved after every sweep.
\item  INITIAL\_N\_NODES [integer number]: Determines the default number of nodes in the initial graph.
\item  INITIAL\_N\_LINKS [integer number]: Determines the default number of links in the initial graph.
\item  NO\_DRAND48: Add this definition if the pseudo-random number function drand48() is not defined on a computer where the program is going to be compiled. In that case the corresponding built-in function generating pseudo-random numbers will be used.
\end{itemize}
An example of the {\tt def.h} file which can be used to generate 100 simple graphs from the canonical ensemble with weight function $p(q)=1/(q+1)$ is:
\begin{lstlisting}
#define ENSEMBLE		2
#define GRAPH_TYPE		3
#define SAVE_FORMAT		3
#define WEIGHT_FUNCTION		1.0/(q+1.0)
#define NV			3000
#define NL			3000
#define SWEEP			5000
#define THERM			100
#define GRAPHS			100
#define INITIAL_N_NODES		100
#define INITIAL_N_LINKS		100
\end{lstlisting}
The choice of ensemble, graphs type, limits for maximal number of nodes and edges as well as the weight function cannot be changed without program re-compilation. The other parameters like input/output format, simulation length etc. can be treated as defaults, since they can be overridden from the command line while starting the program. To make program compilation as easy as possible a {\tt Makefile} is attached. Therefore if one has {\em make} installed, the compilation can be started by issuing the {\em make} command. The resulting executable is called {\tt graphgen.exe}. Every time the file {\tt def.h} is modified, a re-compilation is required before changes take effect.

\subsection{\label{exec}Execution}

To execute the program, type in the command line:\\
{\em graphgen.exe} [options]\\
where [options] can be one or more from the following list:
\begin{itemize} 
\item {\it -h}: \ Help, i.e., print the list off all possible command line options.
\item {\it -n} [integer number]: \ Number of nodes in the initial graph. This number is read from the input file if given.
\item {\it -l} [integer number]: \ Number of links in the initial graph. This number is read from the input file if given.
\item {\it -i} [inputfile]: \ The name of the file with the initial graph. If there is more than one graph saved in the file, only the first is used. If no input file is specified a random graph is generated as the initial graph.
\item {\it-if} [1, 2, or 3]: \ Input file format. Use $1$ for full adjacency matrix format, $2$ for short adjacency matrix format, and $3$ for node degrees format (the details are given below).
\item {\it -o} outputfile: \ Name of the file to which generated graphs are saved. If no file is specified, the program uses standard output.
\item {\it -of} [1, 2, or 3]: \ Output file format (the numbers have the same meaning as for the load format).
\item {\it -r} [any long integer number]: \ Number used to initialize the pseudo-random number generator.
\item {\it -g} GRAPHS: \ Number of graphs to be generated.
\item {\it -s} SWEEP: \ Length of elementary sweep (i.e., number of elementary transformation attempts, see description in subsection \ref{compilation}).
\item {\it -t} THERM: \ Number of initial ``thermalization'' sweeps (see description in subsection \ref{compilation}).
\end{itemize}
For example to generate 100 graphs and save their adjacency matrices to file {\tt graphs.dat} type:\\
{\em graphgen.exe -g 100 -of 1 -o graphs.dat}\\

\subsection{\label{save_load_format}Output data file}
The result of a single program run is the list of generated graphs printed or saved to a file (in turn without empty lines in between). The graphs can be saved in one of three possible formats. In each format the first two lines contain information about the actual number of nodes $nv$ and the number of links $nl$ in the graph. After these two lines the proper information about the graph structure is saved. 

Using the first format, the graph structure is written as an adjacency matrix. Each line contains one row of the matrix. Matrix elements are separated by spaces. For example the output file for the graph in fig. \ref{example_graph} has the form:
\begin{lstlisting}
#nv= 5
#nl= 4
0 0 0 0 1
0 0 0 0 0
0 0 2 0 1
0 0 0 0 2
1 0 1 2 0
\end{lstlisting}
In the second format, only non-zero adjacency matrix elements are saved. Each line in the output file contains information about position (row and column) and value of one non-zero matrix element. Because of the symmetry, it is enough to save information about the upper triangle of the matrix (column $\geq$ row). Thus the graph in fig. \ref{example_graph} would be saved as:
\begin{lstlisting}
#nv= 5
#nl= 4
0 4 1
2 2 2
2 4 1
3 4 2
\end{lstlisting}
If one uses the third format, only nodes degrees are saved. Usually this does not preserve the whole information required to reconstruct the graph but it may be useful, e.g., to construct histograms giving the degree distribution $\pi(q)$. Each line of the output file contains the order of one graph vertex. For the graph in fig. \ref{example_graph} it is:
\begin{lstlisting}
#nv= 5
#nl= 4
0
1
3
4
2
\end{lstlisting}
The same formats are used by the program to load the initial graph from a file. 

\subsection{\label{deg_comp}Compilation of degdist}
We now come to {\em degdist}. This is an independent program, which makes use of some functions defined in source files {\tt init.c}, {\tt links.c} and {\tt sweep.c} described in previous sections. These files are included during the compilation by means of the \lstinline!#include! directive. Thus the program can be compiled as a single file, without any special arrangements. One can also use attached {\tt Makefile} and issue the command {\em make degdist}, which will generate the {\tt degdist.exe} executable file.

Constants used in the program have the same meaning as it was already described. As a default all constants are defined in {\tt degdist.c}, but for convenience there is an option to use the definition from {\tt def.h} file, exactly as it was in the {\em graphgen} program. The only one additional constant:
\begin{itemize}
\item HIST "name"
\end{itemize}
defines the name of the output file into which the histogram of the measured degree distribution is  saved.

An example of constants definition is given below:
\begin{lstlisting}
#define ENSEMBLE 2
#define GRAPH_TYPE 3
#define RATIO_WEIGHT_FUNCTION (q<1)?1e+20:(q*(q+1.)/(q+3.))	
#define SWEEP 500
#define THERM 10000
#define GRAPHS 100000
#define INITIAL_N_NODES 100
#define INITIAL_N_LINKS 100
#define NV 30000
#define NL 30000
#define HIST "test.dat"
#define NO_DRAND48
\end{lstlisting}
This allows to generate $10^5$ pseudographs from the canonical ensemble with $N=100,L=100$ and Barab\'{a}si-Albert degree distribution \cite{ref:bamodel}:
\begin{equation}
	\pi(q) = \frac{4}{q(q+1)(q+2)}
\end{equation}
which leads to $p(q)=4q!/(q(q+1)(q+2))$ and $p(q+1)/p(q)$ as given by RATIO\_WEIGHT\_FUNCTION.
Each graph is generated from the previous one after 500 attempted rewirings. The measured histogram of degree distribution averaged over the generated sample of the canonical ensemble is saved into
{\tt test.dat} file. One can check that this agrees well with the theoretical distribution $\pi(q)$ up to finite-size corrections (cut-off).

\subsection{\label{deg_exec} Execution and output data format of degdist}
After compilation the program {\em degdist} can be executed simply from the command line without any arguments. For parameters given above, the running time is less than one minute on a modern PC. The result of a single run is one data file. Each line consists of three columns separated by tabulators: $q,\pi(q),\Delta\pi(q)$. Here $\pi(q)$ is estimated from measurements of the averaged degree distribution while $\Delta\pi(q)$ gives a rough estimation of the statistical error for this quantity and a given degree $q$. A typical set of data is presented below:

\begin{lstlisting}
1	0.65392		0.00026
2	0.167699	0.00013
3	0.0686797	8.3e-005
4	0.0352373	5.9e-005
5	0.0205495	4.5e-005
6	0.0131624	3.6e-005
7	0.0089971	3e-005
8	0.006363	2.5e-005
...
\end{lstlisting}
where $\dots$ stands for the rest of the file. The $\pi(q)$ given in the second column are normalized such that $\sum_q \pi(q) = 1$. The program can also be compiled with constant {\bf GRAPHS} set to {\bf 1} which means that only one graph is generated and $\pi(q)$ is the degree distribution for this particular graph.

\section{Acknowledgements}
This work was partially supported by Marie Curie Host Fellowship HPMD-CT-2001-00108,
Polish State Committee for Scientific Research (KBN) grant 2P03B-08225 (2003-2006)
and by EU IST Center of Excellence "COPIRA".

\section{Test run}
The program {\em graphgen} has been tested for a number of systems. As an example the results of simulations of a canonical ensemble of pseudographs with $N=3$ nodes, $L=3$ links and the weight function $p(q)=q!\,(q+1)^{-0.5}$ are shown in table \ref{tab2}. The number of $10^7$ graphs have been generated (with \lstinline!THERM=100! and \lstinline!SWEEP=50!). The comparison of graph frequencies calculated theoretically with those generated by the program shows perfect agreement.

The program package contains the example input file {\tt in\_graph.dat} and the example of output file {\tt o\_graph.dat}. In the input file a graph with $N=10,L=50$ is saved in the adjacency matrix format. The output file consists of a list of 20 graphs, saved in the short adjacency matrix format, generated by the following command:

{\em graphgen.exe -g 20 -i in\_graph.dat -if 1 -f o\_graphs.dat -of 2}

The program {\em degdist} has also been tested carefully. The file {\tt test.dat}
contains the degree distribution generated for the set-up given in section \ref{deg_comp} as an example. This was done by compiling and executing {\em degdist.exe} from command line without any arguments.

\newpage
\section*{Figure Captions}

\newpage
\section*{Tables}
\newcommand{\w}[1]{\includegraphics[height=1.9cm]{#1}}
\begin{table}[h]
\begin{center}
\caption{\label{probabilties}Number of possible labelings of graphs with $N=3$, $L=2$, from the canonical ensemble, their weights $w(\alpha)$ and normalized probabilities $p(\alpha)$ for graphs occurrence:
$p(\alpha) = w(\alpha)/\sum_\beta w(\beta)$. \label{tab1}}
\begin{tabular}{|c|c|c|c|c|c|c|}
\hline
graph $\alpha$ & \w{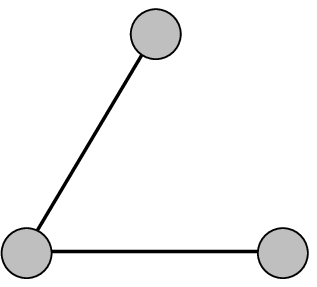} & \w{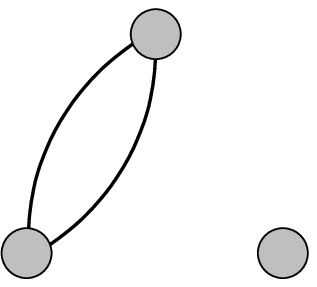} &  \w{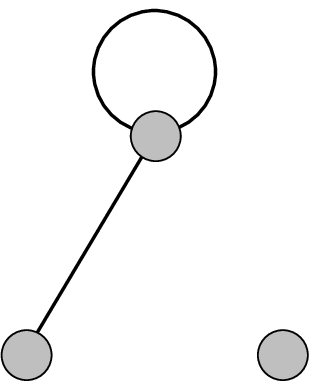} & \w{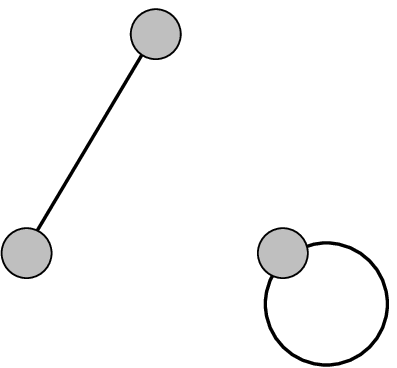} & \w{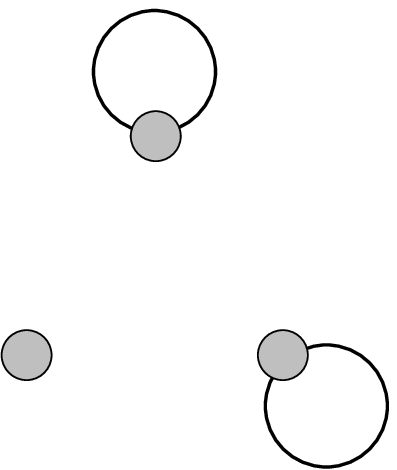} & \w{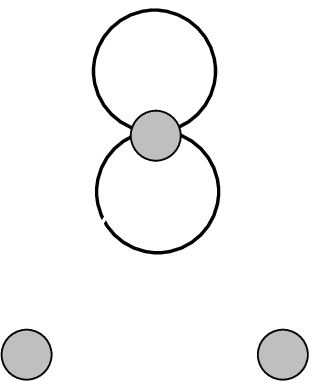}\\
\hline
\#labelings & 72 & 36 & 72 & 36 & 18 & 9\\
$w(\alpha)$ & 1/2 & 1/4 & 1/2 & 1/4 & 1/8 & 1/16 \\
$p(\alpha)$ & 0.2963 & 0.1482 & 0.2963 & 0.1482 & 0.0741 & 0.0370 \\
\hline
\end{tabular}
\end{center}
\end{table}
\vspace{1cm}
\begin{table}[h]
\begin{center}
\caption{Comparison of theoretically calculated frequencies of graph occurrences 
with those generated by the program, for the canonical ensemble with $N=3, L=3$. The weight function is $p(q)=q!\,(q+1)^{-0.5}$. During the simulation $10^7$ graphs were generated (with \lstinline!THERM=100! and \lstinline!SWEEP=50!)\label{tab2}.}
\begin{tabular}{|c|c|c|c|c|c|c|c|}
\hline
\multicolumn{8}{|c|}{\includegraphics[width=13cm]{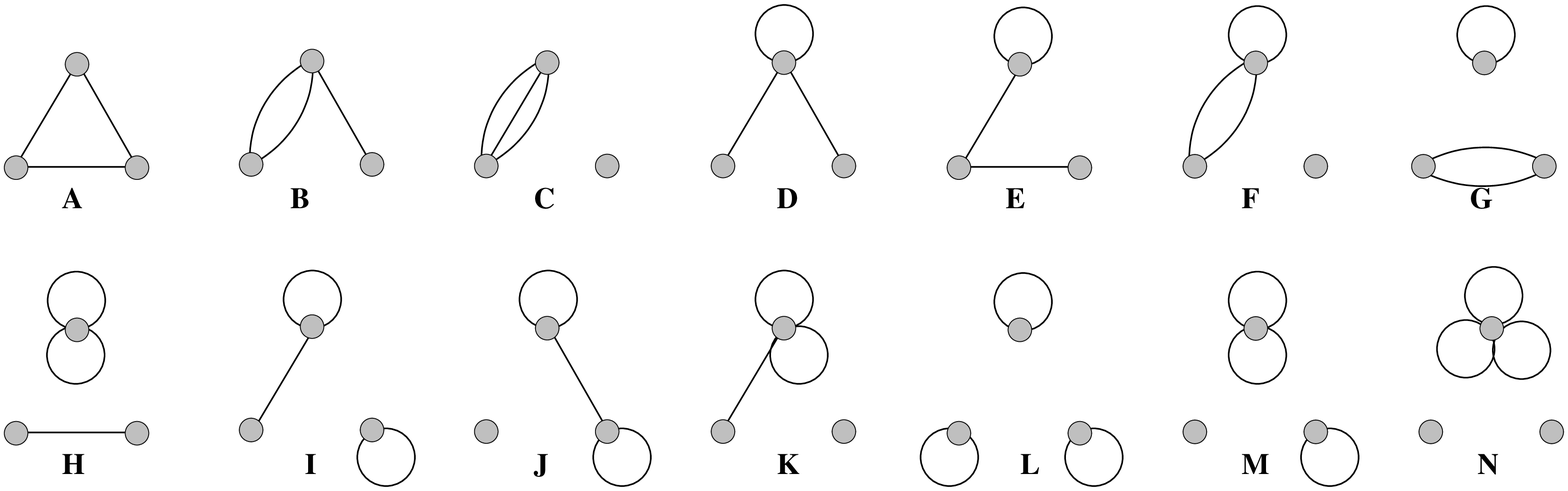}} \\
\hline
graph & A & B & C & D & E & F & G \\
\hline
theor.  & $\frac{p(2)^3}{6}\rightarrow$ & $\frac{p(1)p(2)p(3)}{2}$ & $\frac{p(0)p(3)^2}{12}$ & $\frac{p(1)^2p(4)}{4}$ & $\frac{p(1)p(2)p(3)}{2}$  & $\frac{p(0)p(2)p(4)}{4}$  & $\frac{p(2)^3}{8} \rightarrow$\\
& $0.0142$ & $\rightarrow 0.0675$ & $\rightarrow 0.0414$ & $\rightarrow 0.0740$ & $\rightarrow 0.0675$ & $\rightarrow 0.1708$ & $0.0106$\\
\hline
simul. &  0.0142(1) & 0.0676(1) & 0.0414(1) & 0.0739(1) & 0.0675(1) & 0.1709(1) & 0.0106(1)\\
\hline 
\hline
graph & H & I & J & K & L & M & N \\
\hline
theor.  & $\frac{p(1)^2p(4)}{16}$ & $\frac{p(1)p(2)p(3)}{4}$ & $\frac{p(0)p(3)^2}{8}$ & $\frac{p(0)p(1)p(5)}{8}$ & $\frac{p(2)^3}{48}\rightarrow$ & $\frac{p(0)p(2)p(4)}{16}$ & $\frac{p(0)^2p(6)}{96}$ \\
& $\rightarrow0.0185$ &$\rightarrow 0.0338 $ & $\rightarrow 0.0620$ & $\rightarrow 0.2388$ & $0.0018$ & $\rightarrow 0.0427$ & $\rightarrow 0.1563$\\
\hline
simul. & 0.0184(1) & 0.0338(1) & 0.0621(1) & 0.2389(1) & 0.0018(1) &  0.0427(1)& 0.1562(1)\\
\hline
\end{tabular}
\end{center}
\end{table}


\begin{thebibliography}{99}
\bibitem{ref:ab} R. Albert and A.-L. Barab\'{a}si, Rev. Mod. Phys. {\bf 74}, 47 (2002).
\bibitem{ref:dm} S. N. Dorogovtsev and J. F. F. Mendes, Adv. Phys. {\bf 51}, 1079 (2002).
\bibitem{ref:n1} M. E. J. Newman, SIAM Review {\bf 45}, 167 (2003).

\bibitem{ref:bck} Z. Burda, J. D. Correia and A. Krzywicki, Phys. Rev. E {\bf 64}, 046118 (2001).
\bibitem{ref:bl1} J. Berg and M. L\"{a}ssig, Phys. Rev. Lett. {\bf 89}, 228701 (2002).
\bibitem{ref:dms} S. N. Dorogovtsev, J. F. F. Mendes and A. N. Samukhin, Nucl. Phys. B {\bf 666}, 396 (2003).

\bibitem{ref:bjk1} Z. Burda, J. Jurkiewicz and A. Krzywicki, Physica A {\bf 344}, 56 (2004).
\bibitem{ref:fdpv}  I. Farkas, I. Derenyi, G. Palla and T. Vicsek, Springer Lect. Notes Phys. {\bf 650}, 163 (2004).
\bibitem{ref:pn1} J. Park and M. E. J. Newman, Phys. Rev. E {\bf 70}, 066117 (2004).

\bibitem{ref:kahng} D.-S. Lee, K.-I. Goh, B. Kahng and D. Kim, Nucl. Phys. B {\bf 696}, 351 (2004).
\bibitem{ref:bbw} L. Bogacz, Z. Burda and B. Waclaw, cond-mat/0502124.
\bibitem{ref:er} P. Erd\"os and A. R\'{e}nyi, Publ. Math. Debrecen {\bf 6}, 290 (1959); Publ. Math. Inst. Hung. Acad. Sci. {\bf 5}, 17 (1960).
\bibitem{ref:bb} B. Bollob\'{a}s, {\em Random Graphs}, Academic Press, New York, 1985.
\bibitem{ref:bamodel} R. Albert and A.-L. Barab\'{a}si, Science {\bf 286}, 509 (1999).

\bibitem{ref:bk} Z. Burda and A. Krzywicki, Phys. Rev. E {\bf 67}, 046118 (2003). 
\bibitem{ref:cutoff} M. Bogu\~n\'a, R. Pastor-Satorras and A. Vespignani, Eur. Phys. J. B {\bf 38}, 205 (2004)
\bibitem{ref:cutoff2} S. N. Dorogovtsev, J. F. F. Mendes, A. M. Povolotsky and A. N. Samukhin, cond-mat/0505193.


\bibitem{ref:mc1} D. P. Landau and K. Binder, {\em Monte Carlo Simulations in Statistical Physics}, Cambridge University Press, Cambridge, 2000.
\bibitem{ref:mc2} B. A. Berg, {\em Markov Chain Monte Carlo Simulations and Their Statistical Analysis}, World Scientific, Singapore, 2004.

\bibitem{ref:metrop} N. D. Metropolis, A. Rosenbluth, M. Rosenbluth, A. Teller, E. Teller, J. Chem. Phys. {\bf 21}, 1087 (1953).

\bibitem{ref:maslov} S. Maslov, K. Sneppen and A. Zaliznyak, cond-mat/0205379.

\end{thebibliography}
\end{document}